# Influence of Surface Tension on Nuclear Collective Properties

*Goncharova, N.G.*

M.V.Lomonosov Moscow State University, Department of Physics,119991, Moscow
n.g.goncharova@gmail.com

Rigidities of even-even nuclei were estimated and compared with nuclear charge radii. Correlation of maximal nuclear rigidities with minimal values of $r_0$ parameters was revealed. Influence of effective surface tension on nuclear properties was discussed.

Shell structure in the nucleonic motions has a profound effect on collective properties of the individual nuclei. The competition between the nucleons in the closed shells that promotes the spherical shape of the nucleus and the nucleons in the unfilled shells that tend to polarize the nucleus and bring about nonspherical equilibrium shape is one of the origins in great diversities in nuclear characteristics. This source of differences in nuclear forms and shapes was indicated already at the dawn of nuclear physics [1-3]. These contradictory tendencies reveal themselves as well in the considerable diversities of nuclear responses to excitations, e.g. in great differences in the fragmentations of multipole resonances' strengths. One of nuclear characteristics that reflects the competition between the trends to minimize the nuclear surface and to minimize the Coulomb repulsion is nuclear rigidity. This quality reveals itself in the Hamiltonian of collective vibrations where it determines their potential energy [4]:

$$\hat{H}_{coll.vib.} = \frac{1}{2D}\sum_m |\hat{b}_m|^2 + \frac{C}{2}\sum_m |\hat{a}_m|^2. \quad (1)$$

(Only the most important quadrupole vibrations would be considered)

In (1) $D$ represents the mass transport in the vibrations; C – nuclear rigidity.
The energy of vibration connects both characteristics:

$$\hbar\omega = \hbar\sqrt{C/D}. \quad (2)$$

As was shown in [1], *C* represents the effective surface tension of the nucleus:

$$\tilde{N} = 4R^2\sigma - \frac{3}{10\pi}\frac{e^2 Z^2}{R}. \quad (3)$$

Here σ – coefficient of surface tension, *R* – nuclear radius.

*C* values are connected with the mean squared deformations of nucleus $\beta^2$. In the nuclear state with *N* phonons (e.g. see [4]):

$$\beta_N^2 = \langle N,J,M | \sum_m |\hat{a}_m|^2 | N,J,M \rangle = \frac{\hbar\omega}{2C}(2N+5). \quad (4)$$

For the ground state of a nucleus (*N*=0 state) the rigidity could be estimated as

$$C = \frac{5\hbar\omega_{2^+}}{2\beta_0^2}. \tag{5}$$

In the review [5] the mean squared deformations $\beta_0^2$ are listed for even-even nuclei from $^{10}$Be up to $^{252}$Cf. These values were extracted from measured transitions probabilities $0^+ \to 2^+$. This procedure is not completely model-independent, but the dispersion is not larger than 10-15%. It means that the below shown results of the estimations for $C$ values based on (5) and $\beta_0^2$ from [5] have approximately the same reliability.

Since the rigidities represent the effective surface tensions of nuclei, the systems of nucleons with maximal values of $C$ should have as well the maximal pressure on nuclear matter. In the classical physics of liquids the surface tension of liquid drop, pressure on it, and radii of ellipsoid are connected by the Laplace formula:

$$p = \sigma\left(\frac{1}{R_1} + \frac{1}{R_2}\right). \tag{6}$$

The results for the comparison of the nuclear rigidities for even-even isotopes of the same element and their parameters $r_0$ [6] for charge radii are shown below:

$$r_0 = R_{charge} \cdot A^{-1/3} \tag{7}$$

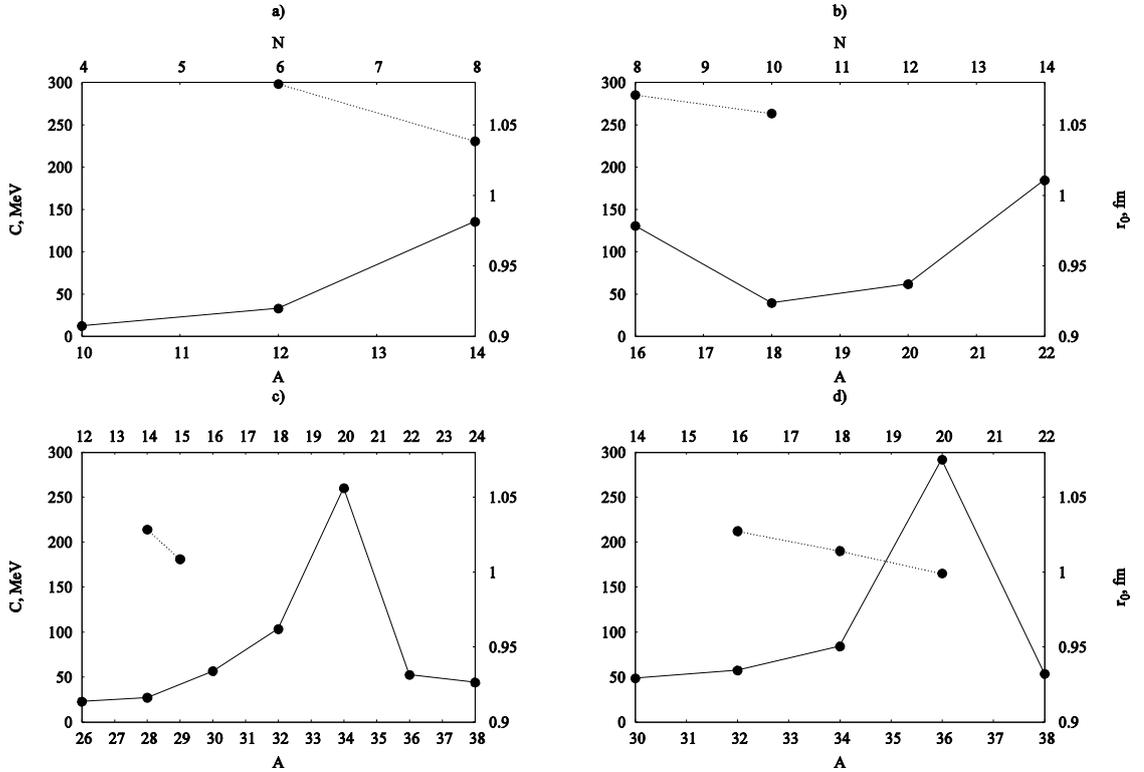

*Fig.1. Rigidities and parameters $r_0$ for Carbon(a), Oxygen(b), Silicon(c) and Sulfur(d) isotopes. Right axes - $r_0$ in Fm; left axes – rigidities C in MeV.*

As can be is seen from the *fig.1*, the maximal values of the nuclear rigidities and consequently of the surface tension correlate with the minimal values of $r_0$. This effect would be demonstrated for almost all investigated even-even nuclei. Since $r_0$ corresponds to charge radii of nuclei under consideration, diminution of $r_0$ marks as well the decrease in the size of proton's well and increase of the charge distributions densities. (The values of nuclear radii used in the *Fig.1* are taken from [6]).

The comparison of rigidities for some light nuclei shows that for the even-even isotopes the rigidity of the nucleus with closed neutron shell is several times larger than for other nuclei of the same element. This effect is especially striking for the even-even calcium isotopes (*Fig.2*).

The $^{48}$Ca nucleus has the maximal rigidity among the light nuclei and simultaneously the minimal $r_0$ parameter [7]. (The systematic investigation of radii for sd-shell nuclei [8] as well reveals the minima of $R_{charge}$ for nuclei with closed neutron shell.)

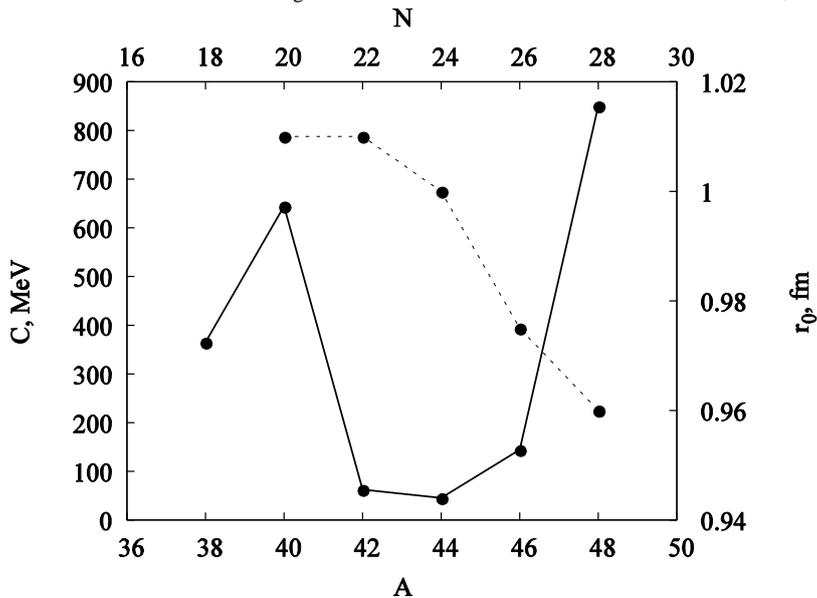

*Fig.2 Rigidities (points connected with solid line) and parameters $r_0$ (points connected with dashed line) for Calcium isotopes.*

The influence of proton shell occupation is demonstrated in the *Fig.3* where the rigidities for the even-even nuclei with Z from 18 up to Z=28 with neutron numbers N=28 are represented.

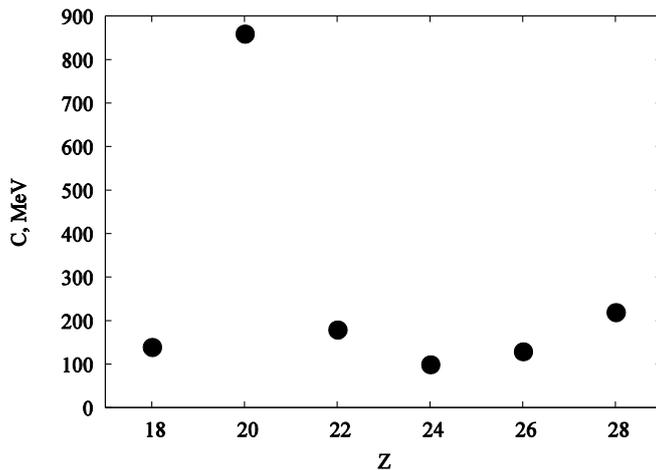

*Fig.3 Rigidities of nuclei with N=28 closed shell:* $^{46}$Ar, $^{48}$Ca, $^{50}$Ti, $^{52}$Cr, $^{54}$Fe, $^{56}$Ni.

The comparison of calculated rigidities for even-even nuclei shows very large diversities in their values. If for Calcium isotopes with A=40 and 48 C>600 MeV, for $^{28}$Si it is less than 20 MeV. The low values of surface tension for $^{28}$Si apparently lead to the deviation of this nucleus from the spherical shape. The effect of this deviation on the structure of E1 resonance was traced in [9].

(It should be mentioned that the calculation of rigidities for $^{40}$Ca and $^{48}$Ca based on a nuclear level scheme was performed in [10]. Obtained values for rigidities are not far from those shown in the Fig.2 but with $C(^{40}Ca) > C(^{48}Ca)$).

A very striking correlation between rigidities and charge radii is revealed from the comparison of these values for Zirconium isotopes (Fig.4). Not only the correlation of $C$ and minimum of $r_0$ for $^{90}$Zr is clearly seen, but a similar effect for $^{96}$Zr as well. Moreover, the value of $r_0(^{96}Zr)$ is less than $r_0(^{90}Zr)$. (The $^{96}$Zr nucleus has two filled neutron shells $(1g_{9/2})_n^{10}(2d_{5/2})_n^6$) near the nuclear surface).

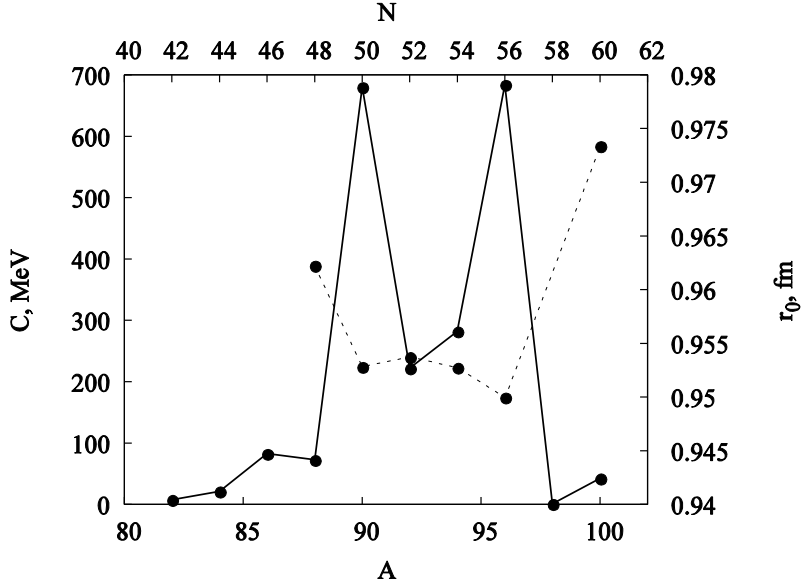

*Fig.4. Rigidities and parameters $r_0$ for Zirconium isotopes*

The investigation of $^{96}$Zr as a "new magic" nucleus with N=56 as magic number was recently performed on the basis of spectroscopic data (see [11] and references there).

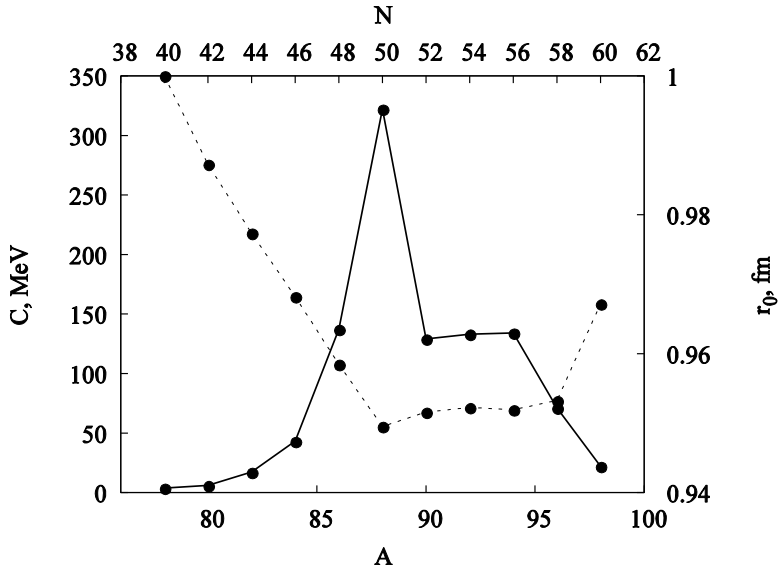

*Fig.5. Rigidities and parameters $r_0$ for Strontium isotopes*

The firmness of "magic" peculiarities in nuclei with neutron numbers N=50 and N=56 are very contrast. The rigidities of nuclei with N=50 are about 3 times higher than those for nuclei of the same isotope with number of neutrons N-2 or N+2. The addition or extraction of a pair of neutrons drastically changes the effective surface tension, exactly as it take place for N=50 in Zr. The comparison of *C* and $r_0$ for nuclei with Z=38 (Strontium isotopes) is shown in the Fig 5. The "magic" number N=50 reveals itself as a prominent peak in rigidities distribution. Similar effect was observed for Z=42 (Molybdenum isotopes) [7].

In both cases (Z=38 and Z=42) the influence of "new magic" number N=56 on the rigidity is visible but much weaker than for Zr isotopes. However the effect of high surface tension on charge radii could be clearly seen.

The correlation between rigidities and nuclear charge radii could be demonstrated for all nuclei with number of neutrons near N=82. In the Fig.'s 6 and 7 it is shown for Ba and Ce even-even isotopes.

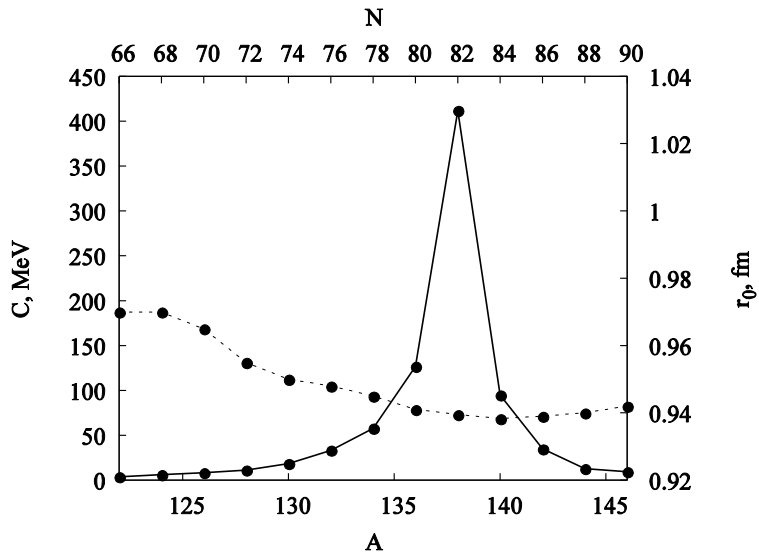

*Fig.6. Rigidities and parameters $r_0$ for Barium isotopes*

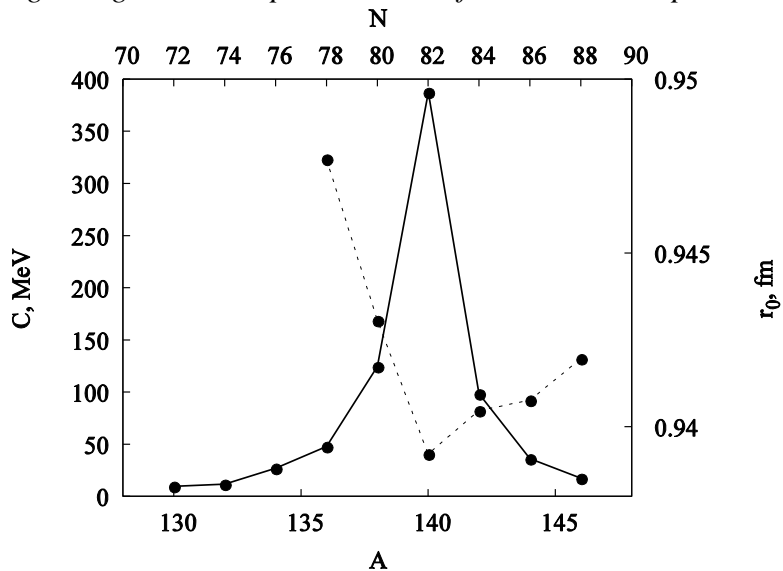

*Fig.7. Rigidities and parameters $r_0$ for Cerium isotopes*

The effect of growing surface tension on nuclear size could be observed in heavy non-magic nuclei as well. The plot of C and $r_0$ parameters for Mercury shown in the Fig.8 reflects the rise of pressure on nuclear matter with adding neutron pairs and manifestation of caused compression on nuclear charge distribution.

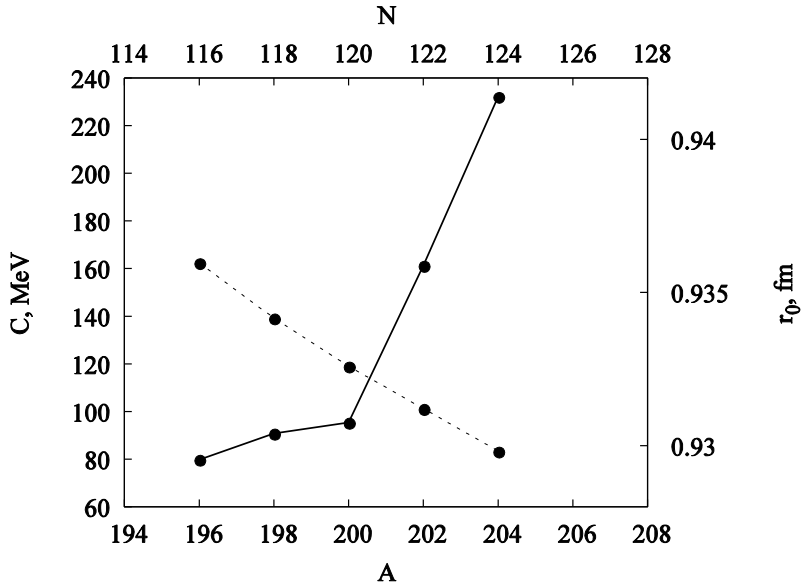

*Fig.8. Rigidities and parameters $r_0$ for Mercury.*

The distribution of rigidities and $r_0$ parameters for even-even isotopes of Lead shown in the Fig.9 demonstrates the maximal values of nuclear rigidities. This is a unique case of a slightly smaller $C$ for a nucleus with magic $N=126$ neutron numbers than for $N=128$. In the performed method of the $C$ estimations it is a consequence of very small parameters of quadrupole deformation $\beta=0.0224$ for $^{210}$Pb in comparison with $^{208}$Pb ($\beta=0.0553$) according to [5].

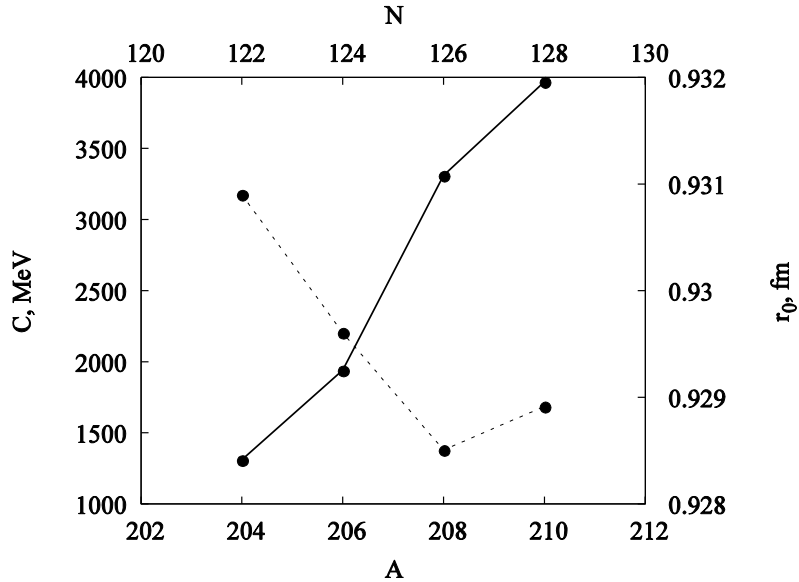

*Fig.9. Rigidities and parameters $r_0$ for Lead.*

CONCLUSIONS

Detected correlations of the nuclear rigidities and nuclear sizes manifest the impact of the surface tension on collective properties of nuclei. The addition of neutron pairs to unclosed neutron shell leads to an increase of surface tension and growing pressure on the nuclear matter.

Since only charge radii of nuclei are systematically explored, the influence of rigidities on nuclear sizes could be properly traced only for proton's distributions in nuclei.

The distributions of neutron matter in nuclei with high rigidities are under investigation.

The author is thankful to Dr.T.Tretyakova and N. Fedorov for helpful discussions.

References

*1. Bohr A.* The Coupling of Nuclear Surface Oscillations to the Motion of Individual Nucleons // Dan.at.Fys.Medd.1952.V.22.#14.P.7-8

*2. Alder K., Bohr A., Huus T., Mottelson B., Winther A.* Study of Nuclear Structure by Electromagnetic Excitations with Accelerated Ions // Rev.Mod.Phys. 1956.V.28. P.433-542

*3. Rainwater J.* Nuclear energy level argument for spheroidal nuclear model //Phys.Rev. 1950.V.79.P.432-435

*4. Eisenberg J., Greiner W.* Nuclear Theory,V.1. 1970 .N.Holl.Amsterdam-London. 445 P.

*5. Raman S., Nestor Jr. C.W., Tikkanen P.* Transition probabilities from the ground to the first–excited $2^+$ state of even-even nuclides // At.Data & Nucl.Data Tables. 2001.V.78,P.1-128

*6. Angeli I.,Marinova K.* Table of experimental nuclear ground state charge radii:An Update //At.Data & Nucl.Data Tables. 2013.V.99 P.69-95

*7. Goncharova N.G., Dolgodvorov A.P., Sergeeva S.I.* Manifestation of shell effects in the collective properties of atomic nuclei// Moscow University Physics Bull.2014.V.69.#3 P.237-242

*8. Blaum K.,Geithner W., Lassen J., Lievens P., Marinova K., Neugart R.*
Nuclear moments and charge radii of argon isotopes between the neutron-shell closures N=20 and N=28 // Nucl.Phys.A .2008.V.799.P.30-45

*9. Goncharova N., Tretyakova T., Fedorov N.* Influence of neutron surface on E1
resonance properties // Nuclear Structure and Related Topics: Book of Abstracts of
the International Conference, Dubna, July 14-18, 2015. JINR Dubna, 2015. P. 37

*10. T.Marumori, S.Suekane, A. Yamamoto.* Nuclear Deformability and Shell Structure //
Progress of Theoretical Physics, 1956. Vol. 16,#4,P.320-340

*11. I.N.Boboshin, B.S.Ishkhanov, V.V.Varlamov*. New data on nuclear subshells
obtained from the analysis of the information from the international database on
nuclear structure ENSDF// Physics of Atomic Nuclei. 2004. V.67. P.1846-1850